\documentclass[10pt]{article}
\pdfoutput=1
\usepackage[utf8]{inputenc}
\usepackage[english]{babel}
\usepackage[pdftex,
            pdfauthor={Giampiero Salvi},
            pdftitle={Developing Acoustic Models for Automatic Speech Recognition in Swedish},
            pdfkeywords={speech recognition hmms}]{hyperref}
\usepackage{graphicx}
\usepackage{url}
\graphicspath{{figures/}}

\title{Developing Acoustic Models for Automatic Speech Recognition in Swedish\footnote{This paper is a summary of the author's Master Thesis that was published in June 1999 on The European Student Journal of Language and Speech (\url{http://www.essex.ac.uk/web-sls/papers/99-01/99-01.html})}}
\author{Giampiero Salvi\\%
\small Kungliga Tekniska Högskolan (KTH)\\%
\small Speech Music and Hearing Department (TMH)\\%
\small SE-100 44 Stockholm, Sweden\\%
\small giampi@speech.kth.se\\%
\small \url{http://www.speech.kth.se/~giampi}}
\date{}

\begin{document}
\maketitle
\begin{abstract}
This paper is concerned with automatic continuous speech recognition using trainable systems. The aim of this work is to build acoustic models for spoken Swedish. This is done employing hidden Markov models and using the SpeechDat database to train their parameters. Acoustic modeling has been worked out at a phonetic level, allowing general speech recognition applications, even though a simplified task (digits and natural number recognition) has been considered for model evaluation. Different kinds of phone models have been tested, including context independent models and two variations of context dependent models. Furthermore many experiments have been done with bigram language models to tune some of the system parameters. System performance over various speaker subsets with different sex, age and dialect has also been examined. Results are compared to previous similar studies showing a remarkable improvement. 
\end{abstract}

\section{Introduction}
The field of speech signal analysis has been in the center of attention for many years because of the many possible applications, but also because, with the many disciplines involved in it, it represents a challenge for many scientists. The applications, related mostly to telecommunication problems, have as a goal the possibility for human beings to exchange information with other human beeings, or with automatic systems, in the most natural and efficient way: speaking. In this context the speech recognition enterprise is probably the most ambitious. Its goal is to build ``intelligent'' machines that can ``hear'' and ``understand'' spoken information, in spite of the natural ambiguity and complexity of natural languages. In thirty years, improvements that could not even be thought of before have been worked out, but still the objective of a robust machine, able to recognize different speakers in different situations, is a very difficult task. The difficulty of the problem increases if we try to build systems for a large set of speakers and for a generic task (large vocabulary).

The thesis summarized in this article describes an attempt to build robust speaker-independent acoustic models for spoken Swedish over the telephone line. A collection of utterances spoken by 1000 speakers (the SpeechDat database) has been used as a statistical base from which models have been developed. The recognition task considered includes a small vocabulary (86 words), but continuous speech is accepted. Furthermore the model structure is chosen with regard to the possibility to apply these models in different contexts and tasks. To evaluate this flexibility the models have been tested on another database from the Waxholm project (vocabulary of 635 words). Different applications are possible for this kind of models: they have already been employed in the recognition part of a complex dialog systems (August project \cite{Gustafson}), or in a speaker verification system (TVIT project \cite{TVIT}). This article contains a documentation of the steps that lead to the creation and development of the acoustic models. 

\section{Speech material}
\label{sec:material}
Model set building and testing are based on a database developed in the SpeechDat project. This database is a subset of the 5000 speakers Swedish database, containing recordings of 1000 subjects. For each speaker (session) a variety of different items are provided for different tasks.

Only a part of them has been used in training and testing the models. 

\subsection{Subjects}
Speakers are selected randomly within a population of interest including all possible types of speakers. Sweden has been divided into seven main dialect areas (``South Swedish'', ``Gotheburg, west and middle Swedish'', ``East, middle Swedish'', ``Swedish as spoken in Gotland'', ``Swedish as spoken in Bergslagen'', ``Swedish as spoken in Norrland'', ``Swedish as spoken in Finland'') . This division does not regards genuine dialects, but rather the spoken language used by most people in the areas defined. 
\subsection{Recordings}
Speech files are recorded through the telephone line and stored in an 8bit, 8kHz, A-law format. For each audio file, an ASCII label file is provided, containing information about sex, age, accent, region, environment, telephone type, and a transcription of the uttered sentence or word. 
\subsection{Items}
The items used in model training contain for each speaker:
\begin{itemize}
\item 9 phonetically rich sentences (S1-S9 in SpeechDat symbols) 
\item 4 phonetically rich words (W1-W4) 
\end{itemize}
while the items used for development tests and evaluation tests are:
\begin{itemize}
\item 1 sequence of 10 isolated digits (B1) 
\item 1 sheet number (5+ digits) (C1) 
\item 1 telephone number (9-11 digits) (C2) 
\item 1 credit card number (16 digits) (C3) 
\item 1 PIN code (6 digits) (C4) 
\item 1 isolated digit (I1) 
\item 1 currency money amount (M1) 
\item 1 natural number (N1) 
\end{itemize}

\subsection{Noise transcriptions}
Noise occurrence is transcribed in the label files for the following cases:
\begin{description}
\item[filled pause] ([fil] in the SpeechDat symbology): is the sound produced in case of hesitation.
\item[speaker noise] ([spk]): every time a speaker produces a sound not directly related to a phoneme generation. 
\item[stationary noise] ([sta]): environmental noise which extends during the whole utterance. 
\item[intermittent noise] ([int]): transient environmental noise extending in a few milliseconds and possibly repeating more than once. 
\item[mispronounced word] (*word) 
\item[unintelligible speech] (**) 
\item[truncation] ($\sim$): $\sim$utterance, utterance$\sim$, $\sim$utterance$\sim$. 
\end{description}
For all these symbols a particular model must be introduced. 

\subsection{Speaker subsets}
In the SpeechDat documentation \cite{speechdat2} a way to design evaluation tests is proposed. For 1000 speaker databases a set of 200 speakers should be reserved for these tests, and the other 800 speakers should be used for \emph{training}. In our case training includes many experiments, and in order to compare them, \emph{development tests} are needed before the final \emph{evaluation test}. For this reason a subset of 50 speakers was extracted from the training set.

In order to maintain the same balance as the full FDB database, speakers for each subset are selected using a controlled random selection algorithm: speakers are divided into different cells with regard on region and gender. Then an opportune number of speakers is randomly selected from each cell to form the required subset. 

\subsection{Statistics}
Results from model tests are presented as mean error rate figures over speakers in the development (50 speakers) and evaluation (200 speakers) subsets respectively. To give an idea of the consistency of these results, some statistics on the database are given in this section.

In Table~\ref{tab:dbstats}, speakers are divided according to gender and age, while in Table~\ref{tab:dialect} the distinction is based on dialect regions.

These distinctions will be particularly useful in Section~\ref{sec:perspeaker}, where per speaker results are discussed. Tables show how the balance in the full database is preserved in each subset. They also show how some groups of speakers are not well represented in the evaluation subset. For example no Finnish speakers nor speakers from Gotland are in this subset. Furthermore speakers from Bergslagen do not represent a good statistical base, a fact to consider when discussing results. The same can be said about young and old speakers in the age distinction.

To give an idea of the consistency of results, in Section~\ref{sec:perspeaker} the standard deviation is reported in addition to the mean values in the case of results for speaker subsets.

\begin{table}
\centering
\begin{tabular}{|c|c|c|c|c|c|}\hline\hline
 Sex & Age & Train & Development & Evaluation & Total \\ \hline 
 F   & young ($<$16) & 21 & 1 & 5 & 27 \\ 
   & middle & 396 & 26 & 106 & 528 \\
   & old ($>$65) & 16 & 1 & 5 & 22 \\ \hline 
   & tot & 433 & 28 & 116 & 577 \\ \hline 
 M & young ($<$16) & 14 & 1 & 1 & 16 \\
   & middle & 284 & 20 & 80 & 384 \\
   & old ($>$65) & 19 & 1 & 3 & 23 \\ \hline 
   & tot & 317 & 22 & 84 & 423 \\ \hline\hline 
\end{tabular}
\caption{Number of speakers in three age ranges for female (F) speakers and male (M) speakers respectively in the Training, Development and Evaluation subsets}
\label{tab:dbstats}
\end{table}

\begin{table}
\centering
\begin{tabular}{|c|c|c|c|c|}\hline\hline 
 Region & Train & Development & Evaluation & Total \\ \hline 
 Bergslagen & 24 & 3 & 4 & 31 \\
 EastMiddle & 281 & 9 & 85 & 375 \\
 Gothenburg & 131 & 12 & 29 & 172 \\
 Norrland & 164 & 11 & 46 & 221 \\
 South & 112 & 9 & 28 & 149 \\
 Finnish & 3 & 2 & - & 5 \\
 Gotland & 6 & 2 & - & 8 \\
 Other & 29 & 2 & 8 & 39 \\ \hline 
 Total & 750 & 50 & 200 & 1000 \\ \hline\hline
\end{tabular}
\caption{Number of speakers belonging to different dialectal areas in the Training, Development and Evaluation subsets}
\label{tab:dialect}
\end{table}

\section{Model sets}
Model set building is based on hidden Markov models (HMMs). Models are employed for ``target'' speech including up to 46 Swedish phonemes, and for ``non-target'' speech including four noise models, one silence model and a word boundary.

\begin{figure}
\includegraphics[width=\textwidth]{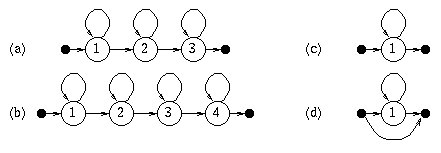}
\caption{Model topology for different applications}
\label{fig:topology}
\end{figure}

\subsection{Target speech}
All phones, except plosives, are modeled by a three emitting state HMM as depicted in Figure~\ref{fig:topology}(a). The choice of topology in HMM applications is often made in the attempt to obtain a good balance between forward transitions and transitions back to the same state. This balance seems to be more important than the correspondence between different states and different parts of the same phoneme realization. This is true for all the steady sounds, such as vowels, fricatives, nasals and so on. The sound produced in the case of plosives (\textbf{B}, \textbf{D}, \textbf{2D}, \textbf{G}, \textbf{K}, \textbf{P}, \textbf{T}, \textbf{2T}), on the other hand, has an important temporal structure, and states in the corresponding HMM should match each different acoustic segment. In a previous work (\cite{melin}), these phonemes have been modeled by a concatenation of two HMMs with three emitting states. In our opinion this method can be inaccurate because of the fast time evolution of plosives. For this reason each plosive is modeled by an HMM with four emitting state (Figure~\ref{fig:topology}(b)). 

\subsection{Non-target speech}
Noise models are: \textbf{extral} (SpeechDat mark: [spk]) for speaker noise such as lips smack, \textbf{Öh} ([fil]) for hesitation between a word and another, \textbf{noise} ([int]) for non speaker intermittent noise typical of the telephone lines. It is not possible to use an HMM to soak up the stationary noise ([sta]) because this disturbance is extended to the whole utterance. The acoustic characteristics of this noise are in part held by the models parameters which are estimated both on ``good'' and noisy files.
 
The \textbf{sil} model is a three state HMM (Figure~\ref{fig:topology}(a)) trained on silence frames of the utterance. In the recognition task it is used at the beginning or at the end of a sentence in the attempt to model the extra time in the recording session over the spoken utterance.

The \textbf{\#} model is employed for word boundaries: it has a symbolic use, representing the boundaries between words at the phone level and allowing different context expansion methods. A second reason for its use is to model the silence between one word and another. In continuous speech these silence segments can be very short. For this reason \textbf{\#} has only one emitting state as depicted in Figure~\ref{fig:topology}(d). This state is tied (shared parameters) with the central state of the \textbf{sil} model. Furthermore a direct transition from the first (non emitting) state to the last (non emitting) state is allowed in the case that words are connected without any pause. Finally the \textbf{garbage}
 model (Figure~\ref{fig:topology}(c)) is created to allow using files containing pronunciation errors (* in SpeechDat), unintelligible speech (**) and truncation ($\sim$). This model is not used during the recognition phase because only ``clean'' files are used in this case. 

\subsection{Retroflex allophones}
During this work the lexicon file has been changed, to include \emph{retroflex allophones } \cite{fant} in a first time considered too rare. For this reason in each experiment two model sets have been tested including (\emph{new lexicon}) or not (\emph{old lexicon}) models for these sounds.

\section{Training}
Training consists of applying an embedded version of the Baum-Welsh algorithm \cite{dtpss}. For every speech file (output sequence) a label file with a phoneme transcription is loaded and used to create an HMM for the whole utterance concatenating all models corresponding to the sequence of labels.

The \textbf{garbage } model is first trained apart on generic speech and then included in the model set.

\subsection{Development tests}
\label{sec:development}
During training development tests are an important tool in the attempt to compare different experiments and to suggest new possible ways of improvement. These experiments consist of a word level recognition on a small subset of speakers reserved for this task. The number of speakers (50) involved in these tests is too low to guarantee a statistical consistency of the results, nevertheless these tests can be used for a comparative evaluation of different model sets. When scoring the system, two alternative parameters are taken into account: correct words and accuracy. Their definition depends on the the algorithm used to align the output of the system to the reference transcription. This algorithm is an optimal string match based on dynamic programming \cite{HTKBook}. Once the optimal alignment has been found, the number of substitution errors (S), deletion errors (D) and insertion errors (I) can be calculated.

Correct words (PC) are then defined as:
\begin{center}
 PC = (N-D-S)/N 100\%
\end{center}
 While the definition for accuracy (A) is:
\begin{center}
 A = (N-D-S-I)/N 100\%
\end{center}
Files chosen for evaluation contains isolated digits, connected digits and natural numbers.

In the following sections development tests results (50 speakers) are presented for different experiments. 

\subsubsection{Monophones}
\begin{figure}
\centering
\includegraphics[width=0.5\textwidth]{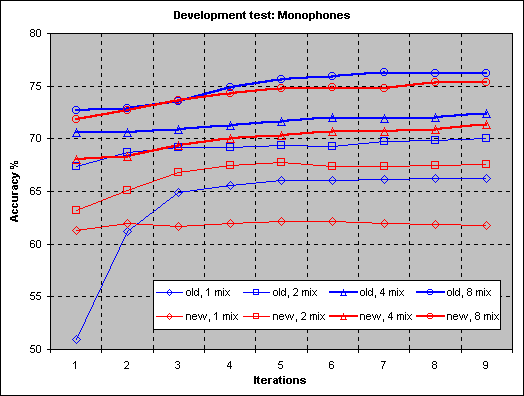}
\caption{Development test (50 speakers). Monophones, 1, 2, 4, 8 Gaussian terms, 
 old lexicon (blue line) and new lexicon (red line)}
\label{fig:monophones}
\end{figure}

Nine iterations of the Baum-Welch re-estimation have been performed. As showed in Figure~\ref{fig:monophones}, recognition performance in terms of word accuracy is improved until the fifth iteration in the case of old lexicon models. Further iterations can be avoided since they don't bring better results. In some cases, as we will see, the performance is even reduced. This is because, with too many iterations, HMM parameters tend to fit too well the training data (and hence the training speaker characteristics) and have no more freedom to generalize to new speakers (evaluation data). In terms of Gaussian parameters it means that the variances tend to be too low (narrow Gaussian shape) to include new speaker variations.

\subsubsection{Triphones}
The construction of context dependent models has been shown to be a good alternative method in the attempt to improving accuracy: Two expansion methods have been tested:
\begin{description}
\item[within-word context expansion] in which phonemes at word boundaries are expanded as diphones. This method is easier to apply in the recognition phase because models created during the network expansion depend only on the words in the dictionary and not on their sequence in the sentence hypothesis. This means that avoiding unseen context dependent models is easy, especially if the words included in the recognition task are also present in the training data.
\item[cross word context expansion] This method results in the generation of a lower number of diphones (only phonemes at sentence boundaries are expanded as diphones). On the other hand, the number of triphone occurrences is increased, rising the context information, and sometimes the number of occurrence for a single model.
\end{description}
Model sets obtained has been subjected to a first iteration of the Baum-Welsh algorithm. For most models the training data was not sufficient, thus a \emph{tree clustering} \cite{synthesis} procedure has been applied. Several threshold values have been used in order to find a good trade-off between the number of states in the model set (model variability) and size of available data (estimation robustness).

\begin{figure}
\begin{minipage}{0.5\textwidth}
\centering
\includegraphics[width=\textwidth]{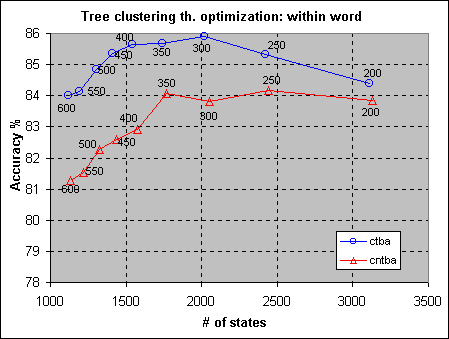} \\ a
\end{minipage}
\begin{minipage}{0.5\textwidth}
\centering
\includegraphics[width=\textwidth]{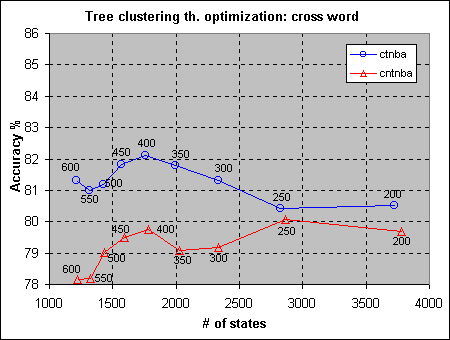} \\ b
\end{minipage}
\caption{a: Tree clustering threshold optimization (Accuracy/\# of states): within word context expanded models, old lexicon (blue line), new lexicon (red line). b: Tree clustering threshold optimization (Accuracy/\# of states): cross word context expanded models, old lexicon (blue line), new lexicon (red line)}
\label{fig:triphones}
\end{figure}

Model sets obtained with different threshold values have been trained separately. Then these models have been tested to find the optimal value for the threshold. Results are shown in Figure~\ref{fig:triphones}.

The optimal model set contains 2020 states in the case of within-word context expansion and the old lexicon (tb) and 2440 states for the new lexicon (ntb); and 1758 and 2862 respectively for old and new lexicon (tnb and ntnb) and cross-word context expansion. Figure~\ref{fig:triphones} also shows that models not including retroflex allophones perform better also in the case of context dependent modeling. The best models have been developed by adding Gaussian mixture terms to the output probability distributions.

\begin{figure}
\begin{minipage}{0.5\textwidth}
\centering
\includegraphics[width=\textwidth]{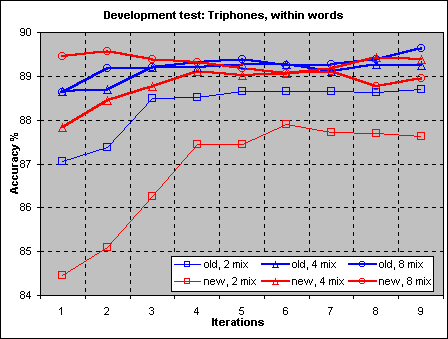} \\ a
\end{minipage}
\begin{minipage}{0.5\textwidth}
\centering
\includegraphics[width=\textwidth]{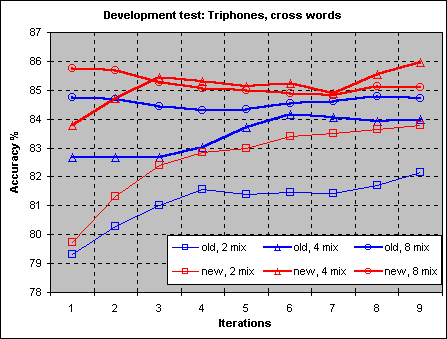} \\ b
\end{minipage}
\caption{a: Accuracy/iterations 1, 2, 4, 8 Gaussian terms: within word context expansion, old lexicon (blue line) and new lexicon (red line). b: Accuracy/iterations 1, 2, 4, 8 Gaussian terms: cross word context expansion, old lexicon (blue line) and new lexicon (red line)}
\label{fig:withincross}
\end{figure}

Results obtained with this method are shown in Figures~\ref{fig:withincross} a and b, respectively for within-word and cross-word context expansion. Figures show nine iterations and models with 2, 4 and 8 Gaussian distributions per mixture. As in the case of monophones, the difference in performance between models including or excluding retroflex allophones is reduced as the mixture size is increased. Within-word models perform better than cross-word models, probably because the number of contexts is lower (6770 models instead of 9681), allowing a more robust parameter estimation. Anyway these results are affected by the specificity of the task. The advantage of using cross word context expansion would probably be higher in a generic speech recognition task in which an higher number of words is involved and sentences are uttered in a more continuous way.

\section{Results}
Model sets selected according to the development tests described in the previous section have been tested on the 200 speakers subset to obtain more reliable results. This section presents overall results and per individual speaker results obtained on the evaluating subset. 
\subsection{Overall results}
From the development tests (Section~\ref{sec:development}) the best models were selected and tested on the evaluation material. Results obtained with these tests are shown in Table~\ref{tab:overall}.
\begin{table}
\centering
\begin{tabular}{|c|c|c|c|c|c|c|c|c|}\hline 
 Experiment & 1 mix & 2 mix & 4 mix & 8 mix & & & & \\ \hline 
        & Corr & Acc & Corr & Acc & Corr & Acc & Corr & Acc \\ \hline 
 mb     & 69.4 & 66.4 & 72.6 & 69.5 & 75.6 & 72.3 & 78.9 & 76.0 \\
 nmb    & 68.1 & 63.1 & 71.5 & 67.9 & 75.1 & 71.3 & 79.1 & 75.5 \\
 ctba   &      &      & 89.5 & 87.4 & 90.7 & 88.5 & 90.8 & 88.6 \\
 cntba  &      &      & 89.1 & 86.4 & 90.3 & 88.1 & 90.5 & 88.3 \\
 ctnba  &      &      & 86.1 & 81.8 & 87.8 & 84.0 & 88.4 & 84.8 \\
 cntnba &      &      & 86.8 & 84.2 & 88.4 & 86.1 & 88.9 & 86.5 \\ \hline 
\end{tabular}
\caption{\textbf{Acc}uracy and \textbf{Corr}ect words for evaluation tests (200 speakers): \textbf{mb} = monophones, old lexicon; \textbf{nmb} = monophones new lexicon; \textbf{ctba} = triphones, within word context expansion, old lexicon; \textbf{cntba} = triphones, within word context expansion, new lexicon; \textbf{ctnba} = triphones, cross word context expansion, old lexicon; \textbf{cntnba} = triphones, cross word context expansion, new lexicon}
\label{tab:overall}
\end{table}
In the table correct words and accuracy are reported for each experiment. The best result (88.6\% of accuracy) is obtained with within-word context expanded models and eight mixture terms. As can be seen in the table, monophone accuracy rises when the number of mixture terms is increased from four to eight. This means that probably better results can be obtained if the number of mixture terms is further increased. In the case of context dependent models the increase of accuracy from four terms models to eight terms models is quite low. Old lexicon models perform better in general than new lexicon models. Models excluding cross-word context information perform better than models including it. 

\subsection{Results per individual speaker}
\label{sec:perspeaker}
SpeechDat database is built on a wide range of speaker characteristics (Section~\ref{sec:material}). For this reason it is interesting to show per speaker results. Often in speaker independent recognition tasks, speakers are divided into ``goats'' and ``sheeps'' depending on results obtained. ``Goats'' are those speakers for which bad results are obtained, while ``sheep'' speakers are well recognized by the system. The definition of the threshold separating these groups is arbitrary and depends on the application.

\begin{figure}
\begin{minipage}{0.5\textwidth}
\centering
\includegraphics[width=\textwidth]{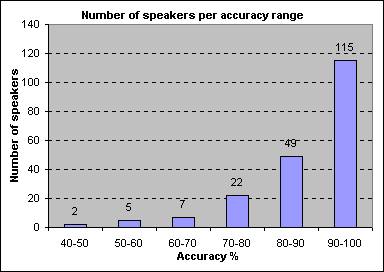} \\ a
\end{minipage}
\begin{minipage}{0.5\textwidth}
\centering
\includegraphics[width=\textwidth]{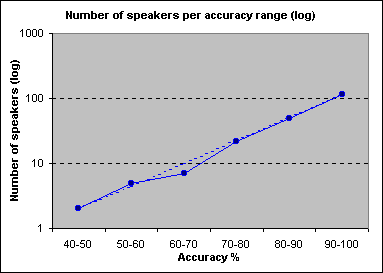} \\ b
\end{minipage}
\caption{a: Number of speakers in 10\% Accuracy ranges. Number of speakers in 10\% Accuracy ranges (log)}
\label{fig:speakers}
\end{figure}
In Figure~\ref{fig:speakers} a, the number of speakers for which results are in ranges of ten percent of accuracy are shown. No speaker in the evaluation subset has results below 40\% of accuracy. If we set the boundary between ``goats'' and ``sheeps'' at 80\% of accuracy, 36 speakers of the 200 in the evaluation subset belong to the ``goats'' group while the other 153 are ``sheeps''. In Figure~\ref{fig:speakers} b, the same data is plotted in a logarithmic scale showing a linear behavior ($y \approx m+ax$). Knowing $m$ and $a$ can be useful to predict results when new speakers are added to the evaluation set, or to evaluate new developments in the systems.

\begin{figure}
\begin{minipage}{0.5\textwidth}
\centering
\includegraphics[width=\textwidth]{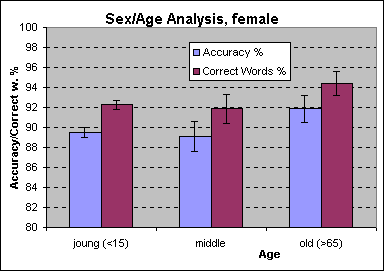} \\ a
\end{minipage}
\begin{minipage}{0.5\textwidth}
\centering
\includegraphics[width=\textwidth]{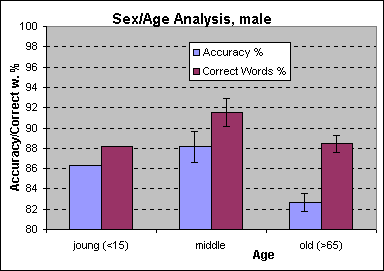} \\ b
\end{minipage}
\caption{Accuracy (left bar) and Correct Words (right bar) for young ($<$16), middle, and old ($>$65) speakers. a) female, b) male.}
\label{fig:gender}
\end{figure}

Figure~\ref{fig:gender} reports results according to sex and age of the speaker, Figure~\ref{fig:gender} a include results on female speakers and Figure~\ref{fig:gender} b on male speakers. In the figure the left bar for each age group displays the accuracy, while the right bar displays the percentage number of correct words. An error bar is also includeded showing the standard deviation in each group (the number of speakers in each group is reported in Table~\ref{tab:dbstats}). Results seem to be sex independent, even though female speakers are better recognized (probably because they are more numerous in the database).

\begin{figure}
\begin{minipage}{0.5\textwidth}
\centering
\includegraphics[width=\textwidth]{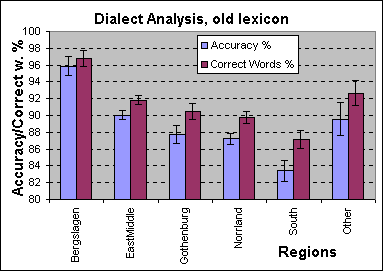} \\ a
\end{minipage}
\begin{minipage}{0.5\textwidth}
\centering
\includegraphics[width=\textwidth]{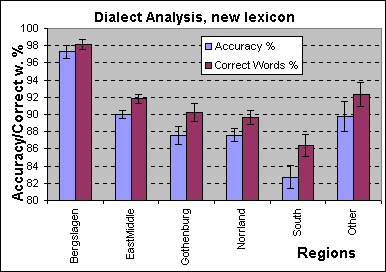} \\ b
\end{minipage}
\caption{Accuracy (left bar) and Correct Words (right bar) for speakers from different regions. Old lexicon (a) and new lexicon (b)}
\label{fig:lexicon}
\end{figure}
Figure~\ref{fig:lexicon} shows results depending on the region the speakers call from. In this case Figure~\ref{fig:lexicon} a shows results obtained with the new lexicon while Figure~\ref{fig:lexicon} b is for old lexicon models results. Speech uttered by speakers from the south of Sweden seems to be more hard to recognize, while speakers from Bergslagen give the best results. An unexpected result refers to speakers from east-middle Sweden, region containing the Stockholm district, and hence the grate part of the Swedish population. In spite of the large amount of training data, for these speakers results are not as high as we would expect. The last comment on Figure~\ref{fig:lexicon} refers to the lexicon. The same trend in per-dialect results is obtained including or excluding retroflex allophones. In the case of southern speakers we would expect lower results for models including the retroflex allophones because speakers living in this region do not make the distinction between normal and retroflex tongue position. In spite of this results are quite similar for southern speakers, new or old lexicon. 
\subsection{Evaluating on Other Databases}
To judge the system performance it would be important to compare results with those obtained with other systems in similar conditions. In our case it is not possible to refer to a previous work on the Swedish SpeechDat database. Comparison is made referring to two experiments which contain substantial differences from the SpeechDat context. For these differences, on the other hand, an evaluation of the model flexibility is possible. The two experiments taken into account refer to the Waxholm project and to the Norwegian SpeechDat project and will be described in the next sections. 
\subsubsection{The Waxholm Database}
The Waxholm database presents many differences if compared to the SpeechDat database.
 
First of all it was developed in the Waxholm project in the attempt to create a dialogue system for information about boat traffic, restaurants and accommodations in the Stockholm archipelago. The corpus of sentences included is hence affected by this task. The number of speakers (mostly men) is low if compared to the SpeechDat collection. 50\% Speakers have an accent typical of the Stockholm area. Speech sampling and labeling are also different (16kHz and phone level by-hand transcriptions). Down-sampling audio files has been necessary because models developed in this work are built for telephone speech. Doing this part of the spectral information in the speech files has been lost.
 
Models are then tested on the same subset of ten speakers used in \cite{kare} on a generic word recognition task, even if model parameters have been tuned in this work with reference to a digit and natural number recognition task. Within-word context expansion models with eight Gaussian distributions per mixture scored 89.9\% of accuracy, while in \cite{kare} 86.6\% of accuracy was reached with sixteen Gaussian distributions triphones.
 
This prove how these models, in spite of the task adopted in this work, can be employed in a more wide range of applications. 

\subsubsection{The Norwegian SpeechDat database}
In Norway similar experiments to those made in this work have been done for Norwegian in the SpeechDat project \cite{speechdat4}. Results are not directly comparable because in our case the same network (loop of words, bigram) has been used for a wide range of different items including for example isolated digits for which a grammar definition allowing only one word (digit) for utterance would be more efficient. However, these results are always similar, even though Norwegian models have been trained on 816 speakers instead of the 750 in our experiments, and the complete database corpus has been used, while only phonetically reach sentences and words have been used in our experiments. 

\section{Conclusions}
Overall results on the evaluation material have shown that models excluding retroflex allophones in general perform better than models including them. This conclusion is surely affected by the task, in fact only a few words included in the recognition task (fyrtio, fjorton, arton, kontokort) contain the allophone 2T, and there is no occurrence of the other allophones. This means that splitting these models in normal and retroflex versions (as in the new lexicon) results only in a lower amount of data for the normal (non retroflex) models mostly used in this task.

Furthermore models including only within-word contexts seem to perform better than models including cross word context information. This result is also affected by the task, first because one of the items included in the evaluating material consists of isolated digits (no cross-word context is available), and second because when uttering digits and natural numbers speakers tend to separate each word to make the sequence clear. In a generic speech recognition task in which a higher number of words is involved and sentences are uttered in a more continuous way, probably the advantage of using cross-word context expansion would be higher.

Per speaker results have shown how models fit quite well to different classes of speakers, with some exceptions. Finally testing models on the Waxholm database has shown the flexibility of these models in spite of the simple task they have been built for. 

\section{Further improvements}
Results for monophone models showed a considerable improvement when passing from four to eight Gaussian terms. For this reason it is likely that further improvements are still possible adding more Gaussian terms. In the case of context dependent models this possibility seems to be more problematic, depending on the fact that the amount of data is not sufficient to train the large number of parameters included in these models. An attempt to reduce this problem could be the use of a model set with a lower number of states respect to the optimal value (tree clustering threshold optimisation) as a base to add Gaussian distributions. The Tree clustering threshold optimisation, indeed, is executed on single distribution models, and there is no reason to think that the number of states in the model set is optimal also when increasing the number of Gaussian parameters.
One experiment in this direction has been tested without good results.

However, the availability of the full 5000 speakers database, will reduce the data scarcity problem allowing the use of more complex models (more Gaussian distributions, lower number of states to be clustered to reduce the number of parameters). Using this database, different strategies will be possible, as for example the creation of two different model sets for female and male speakers, or the creation of dialect dependent models (for example particular models could be built for speakers from the south, which seem to have different characteristics from all the others in Sweden).

Another problem consists in the stationary noise that affects many files in the database. This disturbance, typical of the telephone line, is changing from utterance to utterance affecting the accuracy of acoustic models. A way to reduce the effects of stationary noise could be subtracting the mean energy over each utterance to the mel-cepstral coefficients that constitute the observation sequence to the recognition system. This method, however, requires that the whole utterance is acquired before starting the recognition process, excluding real time applications, in which the speech signal is recorded and analysed on line.

\bibliographystyle{plain}
\bibliography{tesi}

\end{document}